\documentclass[aps,prd,reprint, superscriptaddress,nofootinbib,floatfix]{revtex4-2}

\usepackage{amsmath,amssymb, amsthm,amstext}
\usepackage{natbib}
\usepackage{graphicx}
\usepackage{color}
\usepackage{array, enumerate}
\usepackage{bm}
\usepackage{multirow}
\usepackage[breaklinks,colorlinks,citecolor=blue]{hyperref}
\usepackage{braket}
\usepackage{txfonts}
\usepackage[dvipsnames]{xcolor}

\makeatletter
\def\maketitle{
\@author@finish
\title@column\titleblock@produce
\suppressfloats[t]}
\makeatother
\newcommand{\BH}{{\mbox{\tiny BH}}}
\newcommand{\NS}{{\mbox{\tiny NS}}}
\newcommand{\GB}{{\mbox{\tiny GB}}}
\newcommand{\GR}{{\mbox{\tiny GR}}}
\newcommand{\ST}{{\mbox{\tiny ST}}}
\newcommand{\PPE}{{\mbox{\tiny PPE}}}

\newcommand{\sGB}{{\mbox{\tiny sGB}}}

\begin{document}

\title{Erratum: Constraints on Einstein-dilaton-Gauss-Bonnet gravity from Black Hole-Neutron Star Gravitational Wave Events}
\author{Zhenwei Lyu}
\email{zlyu@uoguelph.ca}
\affiliation{Perimeter Institute for Theoretical Physics, Waterloo, Ontario N2L 2Y5, Canada}
\affiliation{University of Guelph, Guelph, Ontario N1G 2W1, Canada}

\author{Nan Jiang}
\email{nj2nu@virginia.edu}
\affiliation{Department of Physics, University of Virginia, Charlottesville, Virginia 22904, USA}

\author{Kent Yagi}
\email{ky5t@virginia.edu}
\affiliation{Department of Physics, University of Virginia, Charlottesville, Virginia 22904, USA}

\maketitle

\onecolumngrid
The correct version of Eqs.A(8, 10-12), which are coefficients of 0PN, 1PN, 1.5PN and 2PN corrections' should be as below. Note that the corrections only occur for $s_1 s_2$ terms, thus the calculations and figures given in the paper are not affected. This is because when we study the black hole-neutron star system, these $s_1 s_2$ terms will vanish due to neutron star's scalar charge reducing to zero.

\begin{widetext}
\allowdisplaybreaks
\begin{align}
    c_0 & = -\frac{5 \pi}{43008} \frac{(659+728\eta)(m_1^2 s_2 - m_2^2 s_1)^2}{\eta^{5} M^4} - \frac{5 \pi}{16} \frac{s_1 s_2}{\eta^3}\,,\\
\nonumber \\ 
c_1 & = -\frac{5 \pi}{48384} \frac{(m_1^2 s_2 + m_2^2 s_1)^2\,(535+924\eta)}{\eta^{5} M^4}
\nonumber \\
    &\quad - \frac{25 \pi}{576} \frac{(m_1^2 s_2 - m_2^2 s_1)^2}{\eta^{5} M^4} \bigg [ \frac{12497995}{1016064}-\frac{11(m_1-m_2)(m_1^2 s_2 +m_2^2 s_1)}{2 M (m_1^2 s_2 - m_2^2 s_1)} + \frac{15407 \eta}{1440} + \frac{165 \eta^2}{16} \bigg ]\,,\\
\nonumber \\ 
    c_{1.5} & = \frac{\pi^2}{2} \frac{(m_1^2 s_2 - m_2^2 s_1)^2}{\eta^{5} M^4} -\frac{6 \pi^2 s_1 s_2}{\eta^3} -\frac{3 f_3^\GB}{32 \eta}\,,\\
\nonumber \\ 
c_2 & = \frac{5 \pi}{32514048} \frac{1}{\eta^{5} M^5} \bigg [(m_1^5 s_2^2 + m_2^5 s_1^2)(-4341025 + 65553264 \eta - 684432 \eta^2)
\nonumber \\
\nonumber \\
    & \quad + \eta M^2 ( m_1^3 s_2^2 + m_2^3 s_1^2) (20044511 + 65553264 \eta - 
   684432 \eta^2)
\nonumber \\
   & \quad +54 \eta^2 M^5 s_1 s_2 (12952549 - 19310256 \eta - 366128 \eta^2) \bigg ]
-\frac{15  f_4^\GB}{64 \eta}. 
\end{align}
\end{widetext}

\onecolumngrid
\acknowledgements
We'd like to thank Maxence Corman from Perimeter Institute for pointing out this error and all the fruitful discussions with her.

\clearpage

\title{Constraints on Einstein-dilaton-Gauss-Bonnet gravity from Black Hole-Neutron Star Gravitational Wave Events}



\begin{abstract}

Recent gravitational wave observations allow us to probe gravity in the strong and dynamical field regime. In this paper, we focus on testing Einstein-dilaton Gauss-Bonnet gravity which is motivated by string theory. In particular, we use two new neutron star black hole binaries (GW200105 and GW200115). We also consider GW190814 which is consistent with both a binary black hole and a neutron star black hole binary. Adopting the leading post-Newtonian correction and carrying out a Bayesian Markov-chain Monte Carlo analyses, we derive the 90\% credible upper bound on the coupling constant of the theory as $\sqrt{\alpha_\GB}\lesssim 1.33\,\rm km$, whose consistency is checked with an independent Fisher analysis. This bound is stronger than the bound obtained in previous literature by combining selected binary black hole events in GWTC-1 and GWTC-2 catalogs. We also derive a combined bound of $\sqrt{\alpha_\GB}\lesssim 1.18\,\rm km$ by stacking GW200105, GW200115, GW190814, and selected binary black hole events. In order to check the validity of the effect of higher post-Newtonian terms, we derive corrections to the waveform phase up to second post Newtonian order by mapping results in scalar-tensor theories to Einstein-dilaton Gauss-Bonnet gravity. We find that such higher-order terms improve the bounds by $14.5\%$ for GW200105 and $6.9\%$ for GW200115 respectively.

\end{abstract}
 
\maketitle

\section{Introduction} 
Recent updates of the gravitational-wave (GW) catalog (GWTC-3) \cite{LIGOScientific2021djp,LVK2020,aVirgo2014,LIGO2015} reports, in total, 90 gravitational wave events from binary black hole (BBH), binary neutron star (BNS), and neutron star black hole (NSBH) mergers (see \cite{GWTC-1,LIGOScientific:2020ibl,LIGOScientific:2021usb} for the previous catalogs). These events have been used to obtain implications on astrophysics, cosmology, nature of black holes (BHs) and nuclear physics (see studies on e.g. population properties of compact objects \cite{Population2021}, Hubble tension \cite{Hubble2021}, stochastic GW background \cite{stochastic2019}, black hole spectroscopy \cite{qnmkerr2021}, equations of state of neutron stars (NSs) \cite{EoS2018,EoS2019}, and possible mode instabilities driven by NS tidal effects  \cite{pgmode2019,imode2020,imode2021}). 
GW events are also ideal sources to probe strong/dynamical fields of gravity \cite{TheLIGOScientific:2016src,Yunes:2016jcc,graviton2019,Abbott:2018lct,LIGOScientific:2020tif} that are difficult to access through other experiments/observations, including table-top and solar system experiments, or binary pulsar and cosmological observations. For example, they have been used to probe the mass of the graviton \cite{TheLIGOScientific:2016src,Yunes:2016jcc,graviton2019}, scalar-tensor theories (Brans-Dicke theory, those with scalarization phenomena proposed by Damour and Esposito-Far\`ese, screened modified gravity, and the time dependence of the scalar field) \cite{Yunes:2016jcc,Zhao:2019suc,scalar2021},  light axion fields sourced by neutron stars \cite{axion2021}, and dynamical Chern-Simons gravity \cite{Yunes:2016jcc,Nair:2019iur,EdGB_Perkins,Okounkova:2019zjf,Okounkova:2021xjv}). 

Scalar Gauss-Bonnet (sGB) gravity \cite{Nojiri:2005vv,Yagi:2012gp,Antoniou:2017hxj,Antoniou_2018} is another theory beyond General Relativity (GR) that has been studied extensively. In the action, a dynamical scalar field is coupled to a Gauss-Bonnet (GB) invariant (consisting of a certain combination of curvature-squared scalars) with a coupling constant $\alpha_\GB$ that has a dimension of length squared. Depending on what kind of coupling one considers, one recovers a shift-symmetric theory (linear coupling) \cite{Yagi:2011xp,Barausse:2015wia}, Einstein-dilaton Gauss-Bonnet (EdGB) gravity \cite{Kanti:1995vq,Torii:1996yi,Maeda:2009uy,herrerovalea2021shape} (exponential coupling) motivated by string theory and inflation \cite{inflation2020,inflation2021}, and a theory admitting spontaneous scalarization of BHs and NSs (quadratic coupling is an example) \cite{Doneva:2017bvd,Doneva_2018,Silva:2017uqg,Silva:2018qhn}. 

EdGB gravity has been constrained by GWs from BBHs that is summarized in Table \ref{tab:summary}, together with other astrophysical constraints from a BH low-mass x-ray binary (LMXB) and NS observations. The current upper bound on the coupling constant $\sqrt{\alpha_\GB}$ is $\sim 1$km. For example, Perkins et al. \cite{EdGB_Perkins} combined bounds on $\sqrt{\alpha_\GB}$ from 6 selected BBH events from the GW catalogs GWTC-1 and GWTC-2 and found the bound $\sqrt{\alpha_\GB}\lesssim 1.7\,\rm km$. These GW bounds are obtained by taking into account the leading correction to the gravitational waveform phase that enters at $-1$ post-Newtonian (PN) order relative to GR due to the scalar dipole radiation \cite{Yagi:2011xp,Yunes:2016jcc}. Such a correction is derived within the small coupling approximation, where the coupling constant $\alpha_\GB$ is assumed to be much smaller than the characteristic curvature scale of a system (e.g. the mass for a BH) and one keeps only to $\mathcal{O}(\alpha_\GB^2)$. 
Under this approximation, EdGB gravity effectively reduces to shift-symmetric GB gravity with a linear coupling between the scalar field and the GB invariant.

\renewcommand{\arraystretch}{1.2}
\begin{table*}[tb]
\begin{centering}
\begin{tabular}{r|c|c|c|c|c|c}
\hline
\hline
\noalign{\smallskip}
\multirow{2}{*}{}  & \multirow{2}{*}{LMXB} & \multirow{2}{*}{NS} &
 \multicolumn{2}{c|}{GW (BBH)} &  \multicolumn{2}{c}{\textbf{GW (NSBH) (this work)}} \\
\cline{4-7}
&  & & O1--O2 & O1--O3 & \textbf{GW200115} & \textbf{combined}   \\
\hline
$\sqrt{\alpha_\GB}$ [km]& 1.9 \cite{Yagi:2012gp} &1.29 \cite{Saffer:2021gak} & 5.6 \cite{Nair:2019iur}, 1.85~\cite{Yamada:2019zrb}, 4.3~\cite{ppE2018} &  1.7 \cite{EdGB_Perkins}, 4.5 \cite{EdGB_Wang},  (0.4) \cite{EdGB_Wang}  & 1.33 & 1.18 \\
\noalign{\smallskip}
\hline
\hline
\end{tabular}
\end{centering}
\caption{Astrophysical bounds on EdGB gravity. We show bounds from a LMXB, NSs ($\sim 2M_\odot$ NSs), GWs from BBHs, and NSBHs (this work).  The one in brackets comes from GW190814 assuming that it is a BBH, which has some uncertainty. For NSBH, we present the bound from GW200115 and that by combining NSBHs (GW200115, GW200105, and GW190814; assuming the last one as a NSBH is a conservative choice) and BBHs from~\cite{EdGB_Perkins}.
}
\label{tab:summary}
\end{table*}

In this paper, we derive new bounds on EdGB gravity through GWs from NSBH binaries. Some forecasts on constraining the theory with such systems were made in \cite{forecast2020} based on a Fisher analysis. The authors showed that the existing bounds can be improved further for NSBH binaries with a sufficiently small BH mass. We here derive new bounds through a Bayesian analysis using GW200105 and GW200115 \cite{GW200105_0115}. We also consider GW190814, which is consistent with BBH or NSBH, and find bounds on EdGB gravity for the BBH and NSBH assumptions separately.
We perform Bayesian inference to analyze the above events by adopting IMRPhenomXPHM waveform \cite{Pratten:2020ceb,Pratten:2020fqn,Garcia-Quiros:2020qpx} (a phenomenological inspiral-merger-ringdown waveform for precessing BBHs in GR) as our base GR waveform and include EdGB corrections to the inspiral phase. We set a high frequency cutoff as $f_{\rm high} = 0.018/M$ \cite{Abbott_2019inspiral} (for the total mass $M$ in a unit of second) on the strain data, since the EdGB modifications to the waveform within the PN expansion is only valid for the phase at the inspiral stage. 
We also carry out independent Fisher analyses for cross-checking the results.

We improve previous analyses by deriving and including EdGB corrections to the waveform phase to higher PN orders. Recently, Shiralilou et al. \cite{EdGB2021wfm,EdGB2021wfm02} derived the waveform valid to 1PN order higher than the leading tensor/scalar non-dipole and scalar dipole emission respectively. We update this by taking the waveform in scalar-tensor theories (in the Jordan frame) valid to 2PN relative to the leading for each of dipole and non-dipole contributions \cite{EdGB2016wfm}. 
We apply a conformal transformation in scalar-tensor theories to go from the Jordan frame to the Einstein frame, find the mapping between the scalar fields in scalar-tensor theories and EdGB gravity, and use the scalar charges for BHs and NSs in the latter theory. We checked that this correctly reproduces the leading $-1$PN correction in EdGB gravity known previously \cite{Yagi:2011xp,Yunes:2016jcc}. 

We find the following results. First, using the leading EdGB correction to the phase, we find the 90\% credible upper bound on $\sqrt{\alpha_\GB}$ as $\sqrt{\alpha_\GB}\lesssim 1.33\,\rm km$ for GW200115. This bound is stronger than the bound $\sqrt{\alpha_\GB}\lesssim 1.7\,\rm km$ in \cite{EdGB_Perkins} obtained by combining selected BBHs from GWTC-1 and GWTC-2 catalogs. We also derive combined bounds by stacking posterior distributions on $\sqrt{\alpha_\GB}$ from GW200105, GW200115, and GW190814 (conservatively assuming it is a NSBH), and BBHs considered in \cite{EdGB_Perkins}, and find $\sqrt{\alpha_\GB}\lesssim 1.18\,\rm km$. These results are also summarized in Table \ref{tab:summary}. We next study the effect of including higher PN corrections. We find that such corrections do not make a significant difference on the bound on $\sqrt{\alpha_\GB}$ from the case with the leading correction, but improve the bound by  $14.5\%$ for GW200105 and $6.9\%$ for GW200115 respectively. Such a finding is consistent with the analysis in \cite{EdGB_Perkins}.

This paper is organized as follows. We first review EdGB gravity and corrections to the waveform phase in Sec. \ref{sec:EdGB}. We next explain in Sec. \ref{sec:methods} two methods of data analysis adopted in this paper, namely Bayesian inference through Markov-chain Monte Carlo (MCMC) and a Fisher analysis. In Sec. \ref{sec:result}, we present our results and conclude in Sec. \ref{sec:discussion}. We use the convention $G=c=1$ throughout the paper.

\section{Einstein-dilaton Gauss-Bonnet Gravity}
\label{sec:EdGB}

Let us first review EdGB gravity within the context of sGB theory and explain corrections to the gravitational waveform from GR.

\subsection{Theory}

We begin by presenting the action for sGB gravity \cite{Nojiri:2005vv,Yagi:2012gp,Antoniou:2017hxj,Antoniou_2018}:
\begin{align}
    S & = \int d^4x \sqrt{-g}\bigg[\frac{R}{16 \pi} - \frac{1}{2}(\nabla\phi)^2 + \alpha_\GB f(\phi)\mathcal{R}_{\rm GB}^2 \bigg]
+ S_m\,.
\end{align}
Here $g$ is the determinant for the metric $g_{\mu\nu}$, $R$ is the Ricci scalar, $\phi$ is a scalar field, $\alpha_\GB$ is the coupling constant between the scalar field and the metric, $S_m$ is the matter action, and 
\begin{equation}
\mathcal{R}_{\rm GB}^2 = R_{\mu\nu\sigma\rho}R^{\mu\nu\sigma\rho} - 4R_{\mu\nu}R^{\mu\nu} + R^2\,,
\end{equation}
is the GB invariant. $f(\phi)$ is an arbitrary function of the scalar field that determines how it is coupled to the metric. EdGB gravity is realized by choosing $f(\phi) = e^{-\gamma \phi}$ for a constant $\gamma$. As shown in \cite{will2021_01,will2021_02}, this theory can be written in a second-order, hyperbolic form that is well-posed for numerical relativity evolution within a range of parameter space.   

String theory predicts even higher order curvature terms in the action that we do include in the analysis. To justify this and treat the theory as an effective field theory, we work in the small coupling approximation scheme (or reduced-order scheme) where we assume that the GR contribution is dominant and handle EdGB corrections as small perturbations. In particular, we define a dimensionless coupling constant
\begin{equation}
\zeta \equiv \frac{16\pi \alpha_\GB^2}{L^4}\,,
\end{equation} 
where $L$ is the characteristic length of the system and assume $\zeta \ll  1$. This technique has been used to find scalar charges of compact objects \cite{Yagi:2011xp,Yagi:2015oca,Berti:2018cxi}, corrections to the GW phase at the inspiral stage \cite{Yagi:2011xp}, and to carry out numerical simulations of BBH mergers \cite{Okounkova:2020rqw}.

Let us study the theory within the small coupling approximation scheme in more detail.
We perturb field equations  in $\alpha_\GB$ and solve them order by order. Then, $\phi = \mathcal{O}(\alpha_\GB)$ and one can expand $f(\phi)$ in small $\phi$ as:
\begin{equation}
f(\phi) = f(0) + f'(0) \phi + \mathcal{O}(\phi^2)\,.
\end{equation}
The first term is a constant and this does not change the field equations from the GR ones as the GB invariant is a topological term and can be rewritten as a total derivative. Thus, the leading effect comes from the second term where the scalar field is linearly coupled to the GB invariant. For this reason, we consider the following action in this paper:
\begin{align}
\label{eq:action}
    S & = \int d^4x \sqrt{-g}\bigg[\frac{R}{16 \pi} - \frac{1}{2}(\nabla\phi)^2 + \alpha_\GB \phi \mathcal{R}_{\rm GB}^2 \bigg]
+ S_m\,,
\end{align}
where we have absorbed $f'(0)$ into $\alpha_\GB$. 
In this theory, BHs can have non-vanishing scalar charges \cite{Yagi:2011xp,Yagi:2015oca} while NSs do not \cite{Berti:2018cxi}.

Current astrophysical bounds on $\sqrt{\alpha_\GB}$ are summarized in Table \ref{tab:summary}. Besides these, one could use electromagnetic radiation emitted by gas or stars orbiting BHs. For example, simulations of the reflection spectrum of thin accretion disks with present and future X-ray missions show that current missions cannot distinguish BHs in GR and those in sGB gravity, while next-generation missions may be able to distinguish them \cite{Xray2017}. Another possibility is to use Solar System experiments, though they are weaker than the astrophysical bounds in Table \ref{tab:summary} by six orders of magnitude \cite{Solar2007,Yagi:2012gp} as the curvature of spacetime in the vicinity of the Sun is much smaller than that of BHs and NSs.


\subsection{Gravitational Waveforms}

We next find EdGB corrections to the gravitational waveform phase. Given that most of the signal-to-noise ratios (SNRs) for GWs from NSBHs and (small mass) BBHs come from the inspiral portion, we focus on the inspiral stage in our analysis. The leading correction to the phase at the inspiral stage enters at $-1$PN order due to the scalar dipole radiation and was derived in \cite{Yagi:2012gp}. Some of the higher PN corrections were recently derived in \cite{EdGB2021wfm,EdGB2021wfm02}. Here, we identified even higher PN corrections using the waveforms in scalar-tensor theories \cite{EdGB2016wfm} (see Appendix \ref{eqn:coef} for details of the derivation). 

Within the stationary phase approximation \cite{Fisher1994,phaseApprox2009}, the waveform in the Fourier space is given by: 
\begin{equation}
  h(f) = A(f) \exp \left[ i\, \Psi(f) \right]\,,\quad \Psi(f) = \Psi_{\GR}(f) + \delta \Psi(f)\,.      
\end{equation}
Here $A(f)$ is the amplitude, $\Psi_\GR$ is the GR phase, and the EdGB correction to the phase $\delta \Psi$ (up to $\mathcal{O}(\alpha_\GB^2)$) is given in a form 
\begin{equation}\label{eqn:dphiTot}
    \delta\Psi =  \sum_{i} \delta\Psi_{\,i\,\rm PN} = \frac{\alpha_\GB^2}{M^4}\sum_{i} \, c_i\,v^{-5+2i}\,.
\end{equation}
Here $v=(\pi M f)^{1/3}$ is the relative velocity of the binary constituents with GW frequency $f$ and the total mass $M = m_1 + m_2$, where $m_1$ and $m_2$ are the masses of the primary and secondary objects of the system. The coefficients $c_i$ up to 2PN order can be found in Appendix \ref{eqn:coef}. We note that corrections at 1.5PN and 2PN terms contain terms that have not been computed yet and are thus not fully complete.

\section{Data Analysis}\label{sec:methods}
In this paper, we carry out two independent analyses to find constraints on $\sqrt{\alpha_\GB}$. The first method is a MCMC analysis based on Bayesian inference by using the publicly-available GW data. The second method is a simpler Fisher analysis that can be used to obtain rough bounds on $\sqrt{\alpha_\GB}$ to cross check the results from the first method. 

Which GW events shall we consider? 
Since the EdGB corrections to the phase are proportional to $\alpha_\GB^2/M^4$, such corrections become larger for systems with smaller total masses. If the data is consistent with GR, this translates to a stronger bound on EdGB gravity. Furthermore, the leading scalar dipole radiation is proportional to the square of the difference in the scalar charges between two objects. This means that we expect to find stronger bounds on $\sqrt{\alpha_\GB}$ for systems with smaller mass ratios ($q=m_2/m_1 < 1$). For these reasons, we will consider the two NSBH events, GW200105 and GW200115, from O3a, (whose total masses are $10.9 M_\odot$ and $7.1 M_\odot$, and mass ratios are $0.22$ and $0.26$, respectively \cite{GWTC-2, GW200105_0115}). We also employ GW190814 \cite{GW190814} whose mass ratio is small (0.11) and the secondary mass is $m_2\approx 2.6 M_\odot$. The system is consistent with both BBH and NSBH, though the probability of a NS with $2.6\,M_{\odot}$ may be small \cite{GW190814NSBH1,GW190814NSBH2,GW190814NSBH3}. Given the uncertainty in the nature of the secondary object, we consider both possibilities of GW190814 being a BBH and a NSBH. We also use GW151226, a BBH with a relatively small mass, to check our results against those found previously \cite{Nair:2019iur,EdGB_Perkins}.

\subsection{Bayesian Inference}\label{subsec:Bayes}
To unveil the basic information of compact binary systems behind GW events, one usually makes use of a reliable method -- Bayesian inference \cite{Bayes2019,Bayes2021}. According to the Bayes' theorem, a posterior probability $p(\mathcal{\boldsymbol{\vartheta}} | d, \mathcal{H} )$ on parameters $\mathcal{\boldsymbol{\vartheta}}$ from data $d$ under a given hypothesis $\mathcal{H}$ is given by:
\begin{equation}
\label{eq:Bayes_theorem}
p(\mathcal{\boldsymbol{\vartheta}} | d, \mathcal{H} ) =  \frac{p(d | \boldsymbol{\vartheta}, \mathcal{H})\; p(\boldsymbol{\vartheta} | \mathcal{H})}{p(d | \mathcal{H})} = \frac{p(d | \boldsymbol{\vartheta}, \mathcal{H})\; p(\boldsymbol{\vartheta} | \mathcal{H})}{\int d\boldsymbol{\vartheta}\; p(d | \boldsymbol{\vartheta}, \mathcal{H})\;p(\boldsymbol{\vartheta} | \mathcal{H})}.
\end{equation}
Here $p(d | \boldsymbol{\vartheta}, \mathcal{H})$ is the likelihood function while $p(\boldsymbol{\vartheta} | \mathcal{H})$ is the prior on $\mathcal{\boldsymbol{\vartheta}}$.
With a stationary Gaussian noise, the log likelihood function $\log p(d|\boldsymbol{\vartheta}, \mathcal{H})$ 
can be expressed as:
\begin{equation}
  \log p(d|\boldsymbol{\vartheta}, \mathcal{H}) =  \log\bar \alpha -\frac{1}{2} \sum_{k}\left< d_k - h_k(\boldsymbol{\vartheta}) | d_k - h_k(\boldsymbol{\vartheta}) \right>,
\end{equation}
where the index $k$ refers to different detectors and $\log\bar \alpha$ is the normalization factor while $d_k$ and $h_k(\boldsymbol{\vartheta})$ are the data and waveform templates from given detectors. 
The inner product between complex functions $a$ and $b$ is defined as:
\begin{equation}
\label{eq:inner_prod}
  \left<a(t)|b(t)\right> = 2\int_{f_{\rm low}}^{f_{\rm high}} \frac{\tilde{a}^*(f)\tilde{b}(f) + \tilde{a}(f)\tilde{b}^*(f)}{S_n(f)} df. 
\end{equation}
Here $*$ refers to a complex conjugate, $S_n(f)$ is the power spectral density (PSD) of given detectors, $f_{\rm low}$ is the low frequency cutoff of GW data (to be explained later), and $f_{\rm high} = 0.018/M$ \cite{Abbott_2019inspiral} is the approximate maximum frequency at the inspiral stage. Notice that $f_{\rm high}$ is not a fixed number but varies among different MCMC realizations. 

For our analysis, the parameters are those in GR plus the EdGB coupling constant $\sqrt{\alpha_\GB}$: 
\begin{equation}
\boldsymbol{\vartheta} = (\mathcal{M}, q, a_1, a_2, \theta_1,\theta_2, \phi_1, \phi_2, \alpha, \delta, \psi, \iota, \phi_{\rm ref}, t_c, D_L,\sqrt{\alpha_\GB})\,.
\end{equation}
Here $\mathcal{M} = (m_1 m_2)^{3/5} / M^{1/5}$ is the detector frame chirp mass, $q=m_2/m_1 (< 1)$ is the mass ratio, $a_A$ are the dimensionless spin magnitudes while $(\theta_A,\phi_A)$ are the polar and azimuthal angles of the spin angular momentum of the $A$th body,  $(\alpha, \delta)$ are the sky location of the binary (right ascension and declination),  $\psi$ is the polarization angle of GWs with respect to the earth-centered coordinates, $\iota$ is the inclination angle of the binary's orbital angular momentum relative to the detector's line of sight, $\phi_{\rm ref}$ is the reference phase at the reference frequency, $t_c$ is the coalescence time, and $D_L$ is the luminosity distance.

We find posterior distributions on all parameters $\boldsymbol{\vartheta}$ for GW events taken from Gravitational Wave Open Science Center (GWOSC) \cite{GWOSC2021} as follows. We perform MCMC samplings through the PyCBC package \cite{pycbc01,pycbc02} and emcee\_pt sampler \cite{emcee} with 500 walkers and 3 temps. We analyze $32 \,\rm s$ of data for GW200105 and $64\,\rm s$ of data for GW200115. Regarding the low frequency cutoff, we set $f_{\rm low} = 20\,\rm Hz$ except for LIGO Livingston for GW200115, where $f_{\rm low} = 25\,\rm Hz$ was used to avoid some excess noise localized at low frequency \cite{GW200105_0115}.

Regarding priors, we assume a uniform distribution on $\sqrt{\alpha_{\GB}}$ with $[0, 5]\, \rm km$ for GW200105, GW200115 and GW190814 (BBH), and $[0, 15]\, \rm km$ for GW190814 (NSBH). As for spin priors, we adopt isotropic spin distribution on $(\theta_A,\phi_A)$ with a high-spin prior on magnitude, $a_1$ and $a_2 \lesssim 0.99$, for all of the MCMC analyses.

For the base waveform model in GR, we adopt IMRPhenomXPHM (that is also used in \cite{EdGB_Wang}) from LALSimulation package \cite{lalsuite}, which is a phenomenological model in the frequency domain that includes spin precession and higher order multipole radiation modes.  As the $(l,m)=(3,3)$ mode is found to be non-negligible for GW200105, GW200115, and GW190814 \cite{GW200105_0115,GW190814}, we include this mode in these events while we only consider the dominant $(l,m)=(2,2)$ mode for GW151226. We adopt IMRPhenomXPHM model that was constructed for BBHs. As for NSBHs, the tidal effects were found to be negligible \cite{GW200105_0115} for the events considered in this paper, and thus it is safe to adopt the same waveform model.

\subsection{Fisher analysis}\label{subsec:Fisher}
We next explain the second method for the data analysis, namely the Fisher information matrix (FIM) method \cite{Fisher1992,Fisher1994,Fisher1998,Fisher2008}, which is valid when the SNR is large and the noise is stationary and Gaussian.

We begin by expanding the log-likelihood function at the maximum likelihood point $\boldsymbol{\vartheta}^{\rm ML}$ for a given hypothesis $\mathcal{H}$:

\begin{align}\label{eqn:loglikeli}
  \log p(d|\boldsymbol{\vartheta}, \mathcal{H}) & \propto -\frac{1}{2} \sum_{k}\left< d_k - h_k(\boldsymbol{\vartheta}) | d_k - h_k(\boldsymbol{\vartheta}) \right>,
\nonumber\\
  & \propto -\frac{1}{2} \sum_{k}\Gamma^{(k)}_{ij}\, \Delta \boldsymbol{\vartheta}^i\, \Delta\boldsymbol{\vartheta}^j\,,
\end{align}
where $\Delta\boldsymbol{\vartheta}^i = \boldsymbol{\vartheta}^{i,\,\rm ML} - \boldsymbol{\vartheta}^i$ is the error of a given parameter relative to the value at maximum likelihood point and $\Gamma^{(k)}_{ij}$ is the FIM evaluated at the maximum likelihood point $\boldsymbol{\vartheta}^{\rm ML}$:
\begin{equation}
    \Gamma^{(k)}_{ij} = \left\langle\frac{\partial h(\boldsymbol{\vartheta})}{\partial \boldsymbol{\vartheta}^i}\; \bigg| \;\frac{\partial h(\boldsymbol{\vartheta})}{\partial \boldsymbol{\vartheta}^j}\right\rangle\,\bigg|_{\boldsymbol{\vartheta}^{\rm ML}}\,,\qquad \Gamma_{ij} = \sum_{k} \Gamma^{(k)}_{ij}\,,
\end{equation}
where the inner product is given in Eq. \eqref{eq:inner_prod} with the power spectral density $S^{(k)}_n$ for the $k$th detector.
Notice that the elements of FIM are partial derivatives of the waveform template with respect to given parameters. Similar to the Bayseian inference, one can introduce a prior to find the posterior distribution on $\boldsymbol{\vartheta}$. We follow \cite{Berti:2004bd} and impose a Gaussian prior, for simplicity, with a standard deviation $\sigma_{\vartheta^i}^{(0)}$ on each parameter. FIM then becomes 

\begin{equation}
\tilde\Gamma_{ij}=\frac{1}{\left(\sigma_{\theta^i}^{(0)}\right)^2}\delta_{ij}+ \Gamma_{ij}\,.
\end{equation}

The inverse of the FIM is an estimator of the error covariance matrix $\Sigma_{ij}$. The standard error is the square root of the diagonal elements of the covariance matrix. For a given parameter $\boldsymbol{\vartheta}^i$, the standard error can be expressed as: 

\begin{equation}
    \sqrt{\left< (\delta\boldsymbol\theta^i)^2 \right>} = \sqrt{\Sigma_{ii}}\,, \quad \Sigma_{ij} = \left( \tilde \Gamma^{-1} \right)_{ij}\,.
\end{equation}

Regarding the base waveform in GR, we follow \cite{Yunes:2016jcc} and use IMRPhenomD instead of IMRPhenomXPHM that was used for the Bayesian inference analysis (as explained in Sec. \ref{subsec:Bayes}). The former is a simpler version of the latter in the sense that it is valid only for spin-aligned systems (i.e. no spin precession) and includes only the dominant mode. This simplification is justified as we only use the FIM analysis to cross check the results from the Bayesian inference which is more robust. Moreover, Perkins et al. \cite{EdGB_Perkins} showed that the difference in the waveform models between IMRPhenomPv2 (a precessing model similar to IMRPhenomXPHM but only includes the dominant mode) and  IMRPhenomD changes the bound on $\sqrt{\alpha_\GB}$ only by $\sim 20\%$. For simplicity, we use a sky-averaged waveform (and rescale the amplitude so that the SNR matches with the observed one) and the parameters for this second method are as follows:
\begin{equation}
\boldsymbol{\vartheta} = \left(\mathcal{M}, q, a_1, a_2,\phi_{\rm ref}, t_c, D_L,\alpha_\GB^2 \right)\,.
\end{equation}
Notice that we take $\alpha_\GB^2$ as our EdGB parameter instead of $\sqrt{\alpha_\GB}$. This is because the former is what enters directly in the waveform and if one chooses to use the latter, the Fisher matrix becomes singular when we take the fiducial value as $\alpha_\GB = 0$ (for the fiducial values of other parameters, we use those reported by LVC and set $\phi_\mathrm{ref} = t_c = 0$). We impose a Gaussian prior \cite{Berti:2004bd} with the standard deviation of $\sigma_{a_1} = \sigma_{a_2} = 1$ and $\sigma_{\phi_\mathrm{ref}} = \pi$.

\section{RESULTS}\label{sec:result}
\subsection{Leading Correction}

\renewcommand{\arraystretch}{1.2}
\begin{table*}[tb]
\begin{centering}
\begin{tabular}{c|c c c c c|c}
\hline
\hline
\noalign{\smallskip}
\multirow{1}{*}{}  & \multirow{1}{*}{GW200105} & \multirow{1}{*}{GW200115} &
 \multicolumn{2}{c}{GW190814} &   \multirow{1}{*}{GW151226} &  \multirow{2}{*}{combined}  \\
 & NSBH  &NSBH & NSBH & BBH & BBH &    \\
\hline
\multirow{2}{*}{Fisher} & \multirow{2}{*}{$1.55$} & \multirow{2}{*}{$0.91$} & \multirow{2}{*}{$7.39$} &  \multirow{2}{*}{$0.90$} &  \multirow{1}{*}{$4.19$} & \multirow{2}{*}{$0.59$} \\
& & &&  & (2.51 \cite{EdGB_Perkins}) & \\
\hline
\multirow{2}{*}{Bayesian} & \multirow{2}{*}{$1.90$} & \multirow{2}{*}{$1.33$} & \multirow{2}{*}{$2.72$} &  \multirow{1}{*}{$0.37$} &  \multirow{1}{*}{$3.43$} & \multirow{2}{*}{$1.18$} \\
& & & & (0.4 \cite{EdGB_Wang})  & (4.4 \cite{EdGB_Perkins}) & \\
\hline
small coupl. &\multirow{2}{*}{4.40} &\multirow{2}{*}{2.94} & \multirow{2}{*}{11.4}&\multirow{2}{*}{1.27} & \multirow{2}{*}{3.81}&\multirow{2}{*}{---} \\
limit & & & & & &\\
\noalign{\smallskip}
\hline
\hline
\end{tabular}
\end{centering}
\caption{
Constraints on $\sqrt{\alpha_\GB}$ [km]  at $90\%$ credible level with Fisher analysis and Bayesian inference from selected NSBH and BBH events. For GW190814, we consider both NSBH and BBH possibilities due to the uncertainty in the nature of the secondary object. These constraints are derived by using the leading phase correction at $-1$PN order, which are improved by approximately 7--15\% if we include higher PN corrections. Our results for GW190814 (BBH) and GW151226 are consistent with those found in previous work shown in brackets. The last column shows the bound by combining posteriors from GW200105, GW200115, GW190814 (NSBH), and the combined posterior from selected BBHs from GWTC-1 and GWTC-2 catalogs obtained in \cite{EdGB_Perkins}.
The last row shows the upper limits on $\sqrt{\alpha_\GB}$ that is valid within the small coupling approximation (Eq. \eqref{eq:small_coupl}). Observe that all the bounds from the Fisher and Bayesian analyses are within these upper limits, showing the validity of our results.}
\label{tab:result}
\end{table*}

We now present our results. Constraints on $\sqrt{\alpha_\GB}$ from various GW events with Bayesian and Fisher analyses are summarized in Table \ref{tab:result}\footnote{Notice that there are some differences in Bayesian and Fisher analyses, such as the waveform modeling (PhenomXPHM vs IMRPhenomD), sGB parameter ($\sqrt{\alpha}_\sGB$ vs $\alpha_\sGB^2$) and its prior (uniform vs Gaussian). This may explain why Fisher bounds are weaker than the Bayesian ones in some cases.}. Here, we only included the leading $-1$PN correction to the waveform phase. Observe that the bounds from the two analyses for each GW event agree within a factor of $\sim 3$. Since the phase corrections are derived within the small coupling approximation, we need to check whether the bounds presented here satisfies this approximation. Following \cite{EdGB_Perkins}, we require
\begin{equation}
\label{eq:small_coupl}
16\pi \frac{\alpha_\GB^2}{m^4} \leq 0.5\,,
\end{equation}
where $m$ is the smallest length scale in the binary. We choose $m=m_2$ (the mass of the smaller BH) for BBH while $m=m_1$ (the mass of the BH) for NSBH\footnote{For simplicity, we use the mass estimates found by LVC assuming GR while Ref. \cite{EdGB_Perkins} used the median values of the masses from posterior distributions including $\sqrt{\alpha_\GB}$.}. We present in Table \ref{tab:result} the upper limit on $\sqrt{\alpha_\GB}$ that satisfies the above bound. 
Notice that all the Fisher and Bayesian bounds satisfy the small coupling approximation and thus are reliable. Notice also that our Fisher and Bayesian results for GW151226 and GW190814 (BBH) are consistent with those in  \cite{EdGB_Perkins,EdGB_Wang}\footnote{Perhaps a small discrepancy in the results for GW190814 (BBH) is due to the fact that we vary the coalescence time $t_c$ in our Bayesian inference while it seems that Ref. \cite{EdGB_Wang} fixed this parameter (at least the posterior distribution on this parameter is not shown in Appendix A of \cite{EdGB_Wang}).}. Our results are also roughly consistent with the forecast made in \cite{forecast2020} for bounds on $\sqrt{\alpha_\GB}$ with NSBHs derived through a Fisher analysis. For example, the bound for a BH mass of $8M_\odot$ and an SNR of 8 (similar to GW200115 where the BH mass is $5.7M_\odot$ and an SNR of 11.4 \cite{GW200105_0115}) was found to be $\sqrt{\alpha_\GB} \lesssim 0.4$km with advanced LIGO's design sensitivity which has a slightly different shape for the noise curve than that with O3 detectors.

The most stringent constraint comes from GW190814 (BBH) though the event is still consistent with NSBH and thus such a bound may not be robust. 
The reason why the bound on $\sqrt{\alpha_\GB}$ is stronger for BBH than NSBH for GW190814 can be understood as follows. First, notice that the leading correction to the phase is proportional to $(m_1^2 s_2 - m_2^2 s_1)^2/M^4$ (see Eq. \eqref{eq:cm1}). Second, let us consider the case $m_1 \gg m_2$ for simplicity. In this case, we find $c_{-1} \propto 1$ for BBH while $c_{-1} \propto q^4$ for NSBH (the scalar charge $s_2$ is 0 for a NS). Thus, the EdGB correction can be much larger for BBH than NSBH.

\begin{figure}
	\centering
    \includegraphics[width=0.48\textwidth]{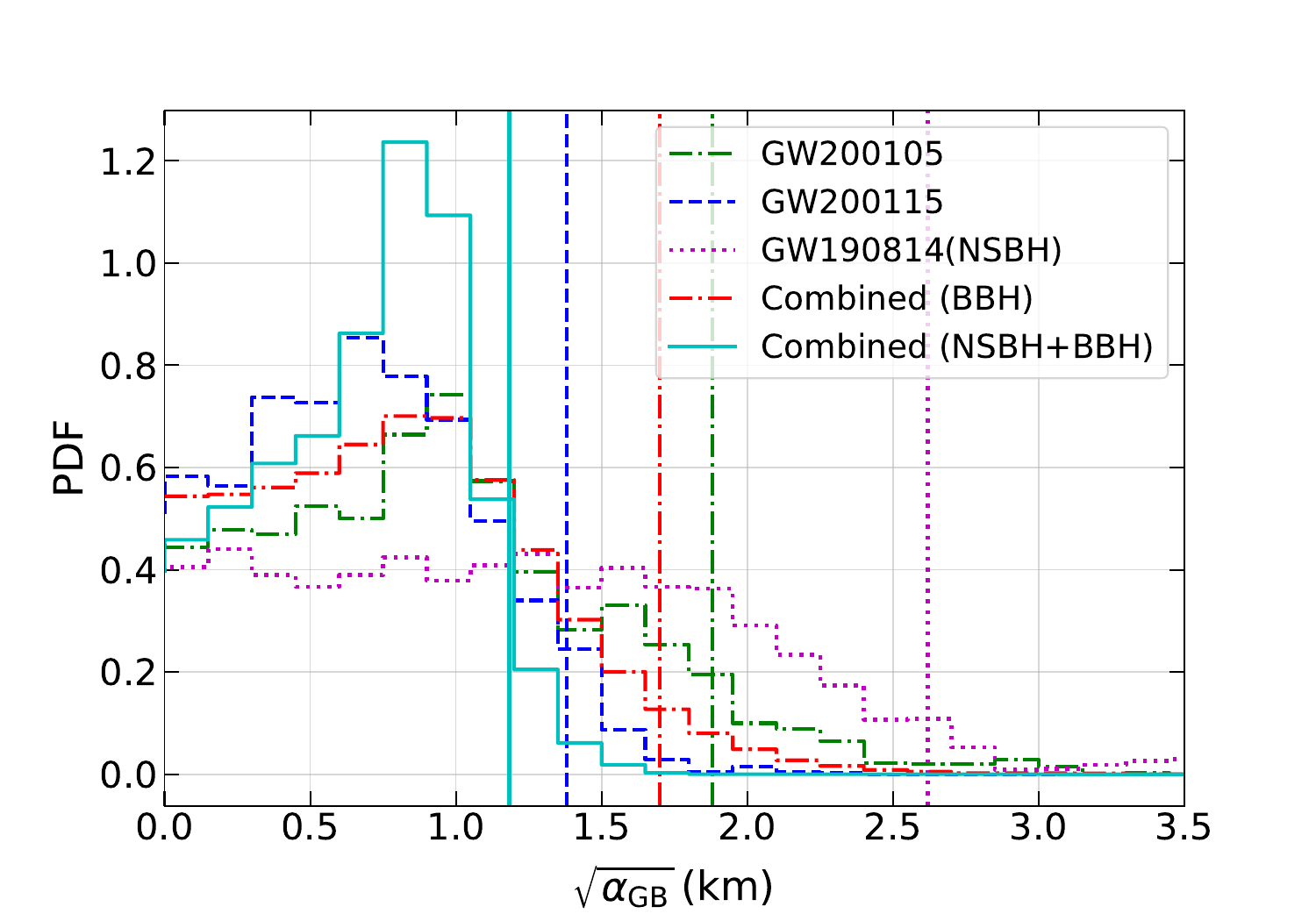}
	\caption{Posterior probability distributions for $\sqrt{\alpha_\GB}$ from selected GW events. We also show an upper bound on $\sqrt{\alpha_\GB}$ at $90\%$ credible level for each event as vertical lines, which indicates the result is consistent with GR. The posteriors are found by including only the leading EdGB correction to the phase at $-1$PN order.} 
	\label{fig:alpha_km}
\end{figure}

Besides constraints from the events GW151226 and GW190814 (BBH) which have already been derived in the previous works \cite{EdGB_Wang,EdGB_Perkins}, we here derived bounds from NSBHs (GW200105, GW200115, and GW190814) for the first time. We present the posterior distributions for $\sqrt{\alpha_\GB}$ for these events in Fig. \ref{fig:alpha_km}. 
The bound from GW200115 is $\sqrt{\alpha_\GB} \lesssim 1.33\,\rm km$, which is stronger than the bound obtained in \cite{EdGB_Perkins} by stacking several BBHs from GWTC-1  and  GWTC-2 catalogs ($\sqrt{\alpha_\GB} \lesssim 1.7\,\rm km$). Observe that the posterior distributions are quite different from Gaussian centered at $\sqrt{\alpha_\GB}=0$, which partially explains the difference between the Fisher and Bayesian results (see also TABLE II, FIG. 2, and FIG. 3 in \cite{EdGB_Perkins}).

Furthermore, we derive combined bounds by multiplying normalized posterior histograms on $\sqrt{\alpha_\GB}$\footnote{This corresponds to the second method discussed in Sec. IIIE of \cite{EdGB_Perkins} for obtaining combined bounds.} from GW200105, GW200115,  GW190814 (with the NSBH assumption that gives us a more conservative bound), and combined BBH bounds in \cite{EdGB_Perkins}. We found a stringent bound of  $\sqrt{\alpha_\GB}\lesssim 1.18\,\rm km$ through the Bayesian analysis as shown in Table \ref{tab:result} and Fig. \ref{fig:alpha_km}.

\begin{figure}
	\centering
    \includegraphics[width=0.495\textwidth]{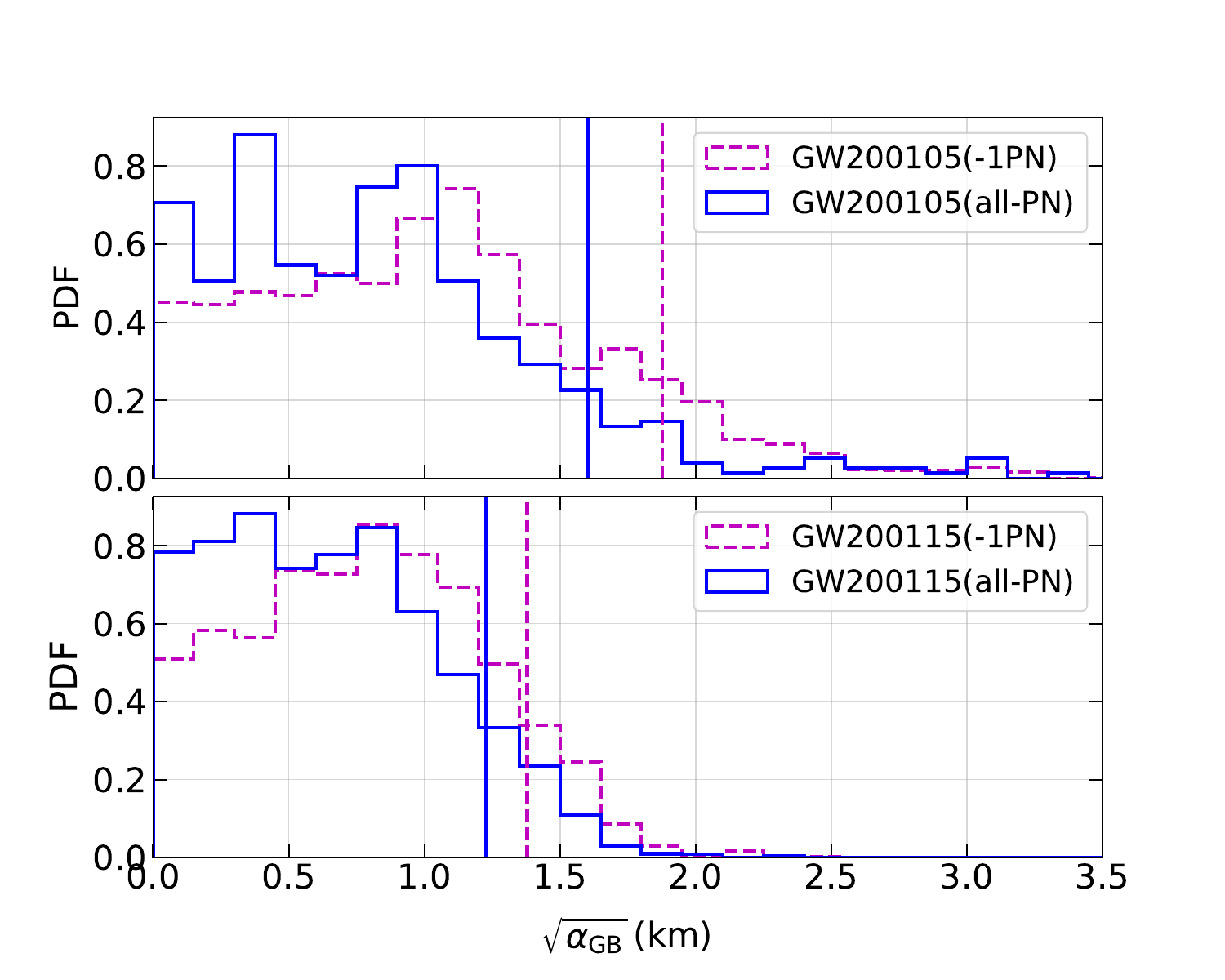}
	\caption{A comparison of the posteriors on $\sqrt{\alpha_\GB}$ from the leading $-1$PN correction and those including higher PN corrections (up to 2PN) for GW200105 (top) and GW200115 (bottom). Observe that the $90\%$ upper bounds on $\sqrt{\alpha}_\GB$ are improved by $14.5\%$ for GW200105 and $6.9\%$ for GW200115 respectively.
} 
	\label{fig:alpha_km_2}
\end{figure}

\begin{figure*}
	\centering
    \includegraphics[width=0.49\textwidth]{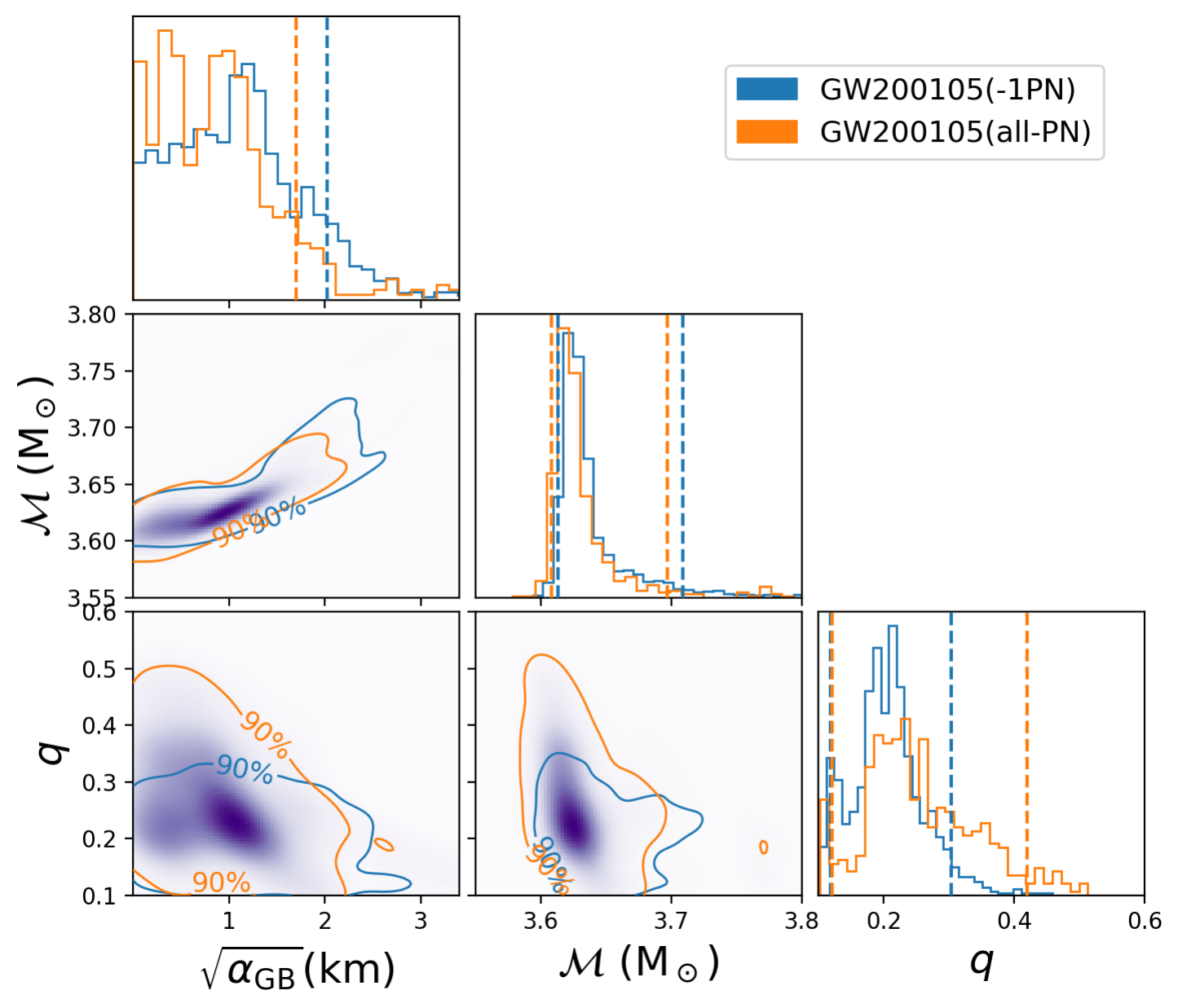}
    \includegraphics[width=0.49\textwidth]{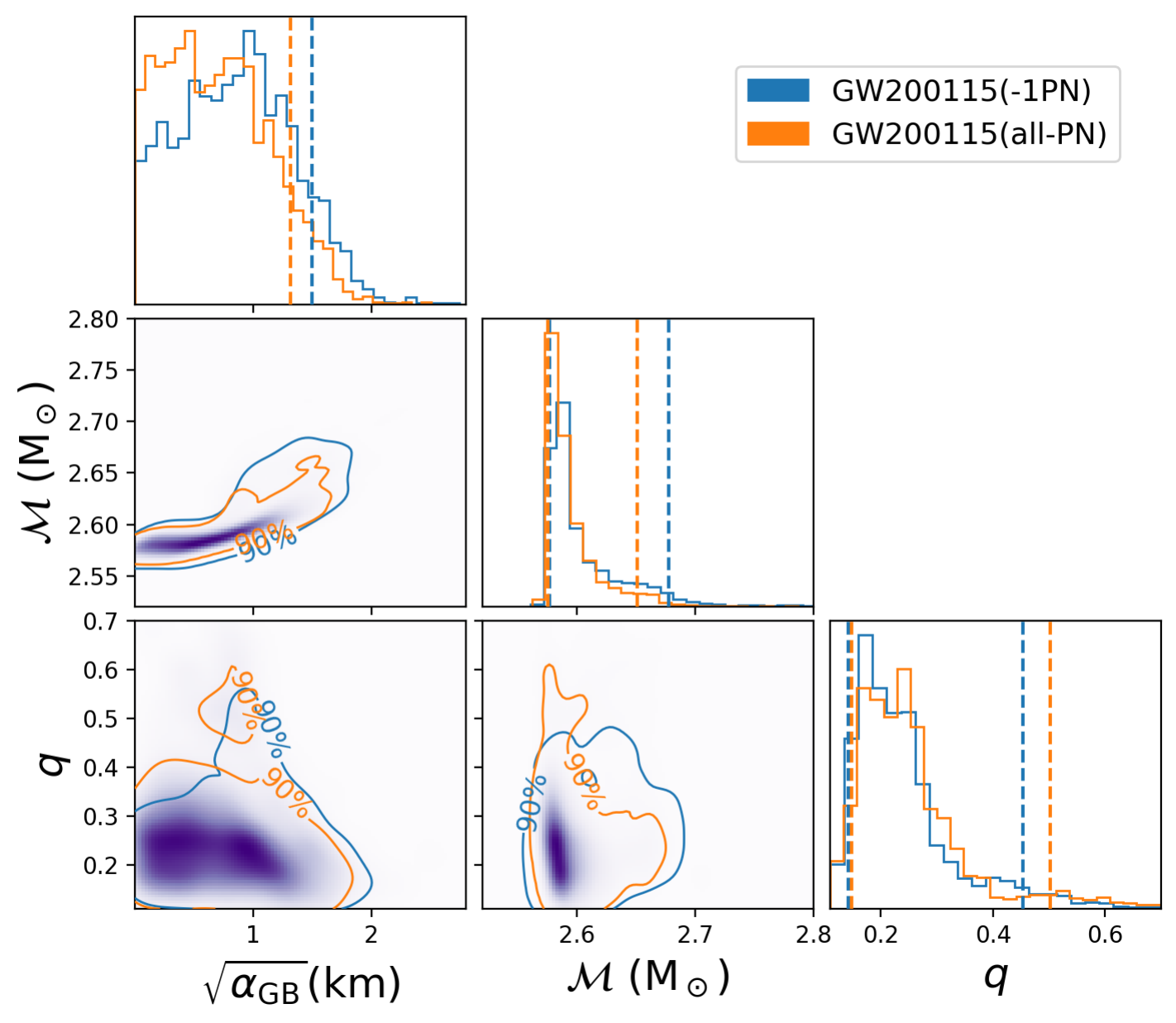}
	\caption{Posterior probability distributions for the EdGB coupling constant $\sqrt{\alpha_\GB}$, the chirp mass $\mathcal{M}$, and the mass ratio $q$ from GW200105 (left) and GW200115 (right). We compare the marginal posterior distributions for the case with the leading EdGB correction at $-1$PN order (blue) and the case including higher PN orders up to 2PN (orange). The purple shaded regions indicate the posterior probabilities of the latter case and the solid lines represent the $90\%$ credible regions for the two cases. The vertical dashed lines show the one-sided $90\%$ confidence interval  for $\sqrt{\alpha_\GB}$ and the two-sided $90\%$ credible intervals for $\mathcal{M}$ and $q$.}
	\label{fig:pos_compare}
\end{figure*}

\subsection{Effects of Higher PN Corrections}

We next study the effect of higher PN corrections to the waveform phase by including PN corrections up to 2PN as presented in Appendix \ref{eqn:coef}. Perkins et al. \cite{EdGB_Perkins} carried out a similar analysis though such higher PN corrections were not available at that time. Thus, the authors considered three different ways to parameterize the unknown 0PN correction (which is 1PN higher than the leading $-1$PN correction) based on the functional forms at 1PN order in GR and the leading $-1$PN EdGB corrections. They then marginalized over such a parameter and concluded that higher PN corrections do not affect the results much and the bounds derived with the leading correction are robust. We check this outcome by using explicit forms of the higher PN corrections in EdGB gravity.

Figure \ref{fig:alpha_km_2} presents posteriors on $\sqrt{\alpha_\GB}$ for GW200105 and GW200115 with and without higher PN corrections, while Fig. \ref{fig:pos_compare} shows corresponding corner plots on $\sqrt{\alpha_\GB}$, $\mathcal{M}$ and $q$. Notice that the inclusion of the higher PN corrections does not affect the posteriors much, especially for GW200115. The 90\% credible upper bound on $\sqrt{\alpha_\GB}$ improves from the case with the leading correction by 14.5\% for GW200105 and 6.9\% for GW200115 respectively. These findings are consistent with those in \cite{EdGB_Perkins} and a very recent work \cite{perkins2022parametrized} that investigated the improvement one obtains when including higher PN order terms.

\section{CONCLUSIONS and DISCUSSION}\label{sec:discussion}

In this paper, we derived bounds on EdGB gravity using GWs from NSBH binaries. Using the leading PN correction, we found $\sqrt{\alpha_\GB}\lesssim 1.33\,\rm km$ as a 90\% credible limit from GW200115, which is stronger than the bound in \cite{EdGB_Perkins} found by combining selected BBHs from GWTC-1 and GWTC-2 catalogs. We also derived combined bounds by stacking posterior distributions on $\sqrt{\alpha_\GB}$ from GW200105, GW200115, GW190814 and the combined posteriors from selected BBHs in \cite{EdGB_Perkins}, and found $\sqrt{\alpha_\GB}\lesssim 1.18\,\rm km$. We further derived higher PN corrections in the waveform phase up to 2PN order from the results in scalar-tensor theories \cite{EdGB2016wfm}. Using these, we improved bounds on $\sqrt{\alpha_\GB}$ for GW200105 and GW200115 from the case with leading PN correction alone by $14.5\%$ and $6.9\%$ respectively.

The analysis carried out here can easily be extended to probe other theories of gravity. We looked at constraining dynamical Chern-Simons gravity \cite{Alexander:2009tp}, which is a parity-violating quadratic gravity whose leading PN correction to the phase is derived in \cite{Yagi:2012vf}. Similar to the case with BBHs \cite{Yunes:2016jcc,Nair:2019iur,EdGB_Perkins}, we were not able to find meaningful bounds that satisfy the small coupling approximation. For future work, one could consider e.g. sGB gravity with the coupling function $f(\phi) \propto \phi^2$ or $f(\phi) \propto 1-e^{-6\phi^2}$ that admits spontaneous scalarization of BHs \cite{Silva:2017uqg,Doneva:2017bvd}.

\acknowledgements
  We thank Huan Yang for useful discussions and thank Reed Essick for helpful comments. Research at Perimeter Institute is supported in part by the Government of Canada through the Department of Innovation, Science and Economic Development Canada and by the Province of Ontario through the Ministry of Colleges and Universities.
 N.J. and K.Y. acknowledge support from the Owens Family Foundation.
 K.Y. acknowledges support from NSF Grant PHY-1806776, NASA Grant 80NSSC20K0523, and a Sloan Foundation. 
 K.Y. would like to also acknowledge support by the COST Action GWverse CA16104 and JSPS KAKENHI Grants No. JP17H06358.
 This material is based upon work supported by NSF's LIGO Laboratory which is a major facility fully funded by the National Science Foundation.
  This research has made use of data, software and/or web tools obtained from the Gravitational Wave Open Science Center (https://www.gw-openscience.org/ ), a service of LIGO Laboratory, the LIGO Scientific Collaboration and the Virgo Collaboration. LIGO Laboratory and Advanced LIGO are funded by the United States National Science Foundation (NSF) as well as the Science and Technology Facilities Council (STFC) of the United Kingdom, the Max-Planck-Society (MPS), and the State of Niedersachsen/Germany for support of the construction of Advanced LIGO and construction and operation of the GEO600 detector. Additional support for Advanced LIGO was provided by the Australian Research Council. Virgo is funded, through the European Gravitational Observatory (EGO), by the French Centre National de Recherche Scientifique (CNRS), the Italian Istituto Nazionale di Fisica Nucleare (INFN) and the Dutch Nikhef, with contributions by institutions from Belgium, Germany, Greece, Hungary, Ireland, Japan, Monaco, Poland, Portugal, Spain.

\appendix

\section{EdGB Corrections to Gravitational Waveforms}\label{eqn:coef}

In this appendix, we explain how to map the waveform (for non-spinning BBHs) in scalar-tensor theories \cite{EdGB2016wfm} to that in EdGB gravity. The former is valid to 2PN order higher than the leading for each of tensor and scalar emission. 

The waveform in scalar-tensor theories is derived in the Jordan frame while EdGB gravity is in the Einstein frame. Therefore, we first turn the former into the Einstein frame. This can be realized by using the mapping provided in Appendix A of \cite{EdGB2016wfm}. After this transformation, the waveform is given in terms of the scalar charge $\alpha_A$ and its derivative $\beta_A$ for the $A$th body. 

The next step is to find these charges in EdGB gravity and substitute this into the waveform. We can compute these following \cite{Julie:2019sab} which uses a slightly different convention for sGB gravity:
\begin{equation}
S  = \frac{1}{16 \pi}\int d^4x \sqrt{-g}\bigg[R - 2(\nabla\varphi)^2 + \alpha_\GB \bar f(\varphi)\mathcal{R}_{\rm GB}^2 \bigg] + S_m\,.
\end{equation} 
One can perform the following rescaling in the scalar field  $\varphi$ and the identification of the function $\bar f(\varphi)$ to recover the action in Eq.~\eqref{eq:action}:
\begin{equation}
\bar f(\varphi) = 2 \sqrt{16\pi} \varphi\,, \quad \varphi = \frac{\sqrt{16\pi}}{2} \phi\,.
\end{equation}
From this, $\alpha_A$ and $\beta_A$ for a non-rotating BH to leading order in $\alpha_\GB$ are given by:
\begin{align}
\alpha_A^\BH =& - \frac{\alpha_\GB \bar f'(\varphi_0)}{ 2 m_{A}^2} =- \frac{ \sqrt{16\pi}\alpha_\GB}{ m_{A}^2}\,, \\
\beta_A^\BH =& -\frac{\alpha_\GB^2  {\bar f'(\varphi_0)}^2}{2 m_A^2}= -\frac{32\pi \alpha_\GB^2 }{m_A^2}\,, 
\end{align}
where $\varphi_0$ is the asymptotic value of the scalar field $\varphi$ at infinity. When substituting these into the waveform expression, the terms with $\beta_A$ enter at $\mathcal{O}(\alpha_\GB^4)$ and are negligible. For $\alpha_A$, we add the spin dependence as:
\begin{equation}
\alpha_A^\BH =- \frac{\sqrt{16\pi} s_A\alpha_\GB}{ m_{A}^2}\,,
\end{equation}
 where the spin dependent factor is given by: \cite{Yunes:2016jcc,Berti:2018cxi}
\begin{equation}
s_A = 2\frac{\sqrt{1-\chi_A^2}-1+\chi_A^2}{\chi_A^2}\,.
\end{equation}
This reduces to $s_A^\BH \to 1$ in the limit $\chi_A \to 0$.
For NSs, $\alpha_A^\NS = \mathcal{O}(\alpha_\GB^3)$ and is negligible while $\beta_A^\NS$ has not been computed. Though we expect the $\alpha_\GB$ dependence to be the same as BH and ignore such terms in the waveform.

Using these charge expressions in the dominant harmonics ($\ell=m=2$) of the waveform and keeping only to $\mathcal{O}(\alpha_\GB^2)$, EdGB corrections to the waveform can be expressed as in Eq. \eqref{eqn:dphiTot} with the coefficients given as follows:
\begin{widetext}
\allowdisplaybreaks
\begin{align}
\label{eq:cm1}
    c_{-1} & =- \frac{5 \pi}{448} \frac{(m_1^2 s_2 - m_2^2 s_1)^2}{\eta^{5} M^4} \,,\\
\nonumber \\ 
\label{eq:c0}
    c_0 & = -\frac{5 \pi}{43008} \frac{(659+728\eta)(m_1^2 s_2 - m_2^2 s_1)^2}{\eta^{5} M^4} + \frac{5 \pi}{8600} \frac{s_1 s_2}{\eta^3}\,,\\
\nonumber \\ 
    c_{0.5} & = \frac{75 \pi^2}{448} \frac{(m_1^2 s_2 - m_2^2 s_1)^2}{\eta^{5} M^4}\,,\\
\nonumber \\ 
    c_1 & = -\frac{5 \pi}{48384} \frac{(m_1^2 s_2 + m_2^2 s_1)^2\,(535+924\eta)}{\eta^{5} M^4}-\frac{5 \pi}{2016}\frac{s_1 s_2 (743 + 924\eta)}{\eta^3}
\nonumber \\
    &\quad - \frac{25 \pi}{576} \frac{(m_1^2 s_2 - m_2^2 s_1)^2}{\eta^{5} M^4} \bigg [ \frac{12497995}{1016064}-\frac{11(m_1-m_2)(m_1^2 s_2 +m_2^2 s_1)}{2 M (m_1^2 s_2 - m_2^2 s_1)} + \frac{15407 \eta}{1440} + \frac{165 \eta^2}{16} \bigg ]\,,\\
\nonumber \\ 
    c_{1.5} & = \frac{\pi^2}{2} \frac{(m_1^2 s_2 - m_2^2 s_1)^2}{\eta^{5} M^4} -\frac{3 f_3^\GB}{32 \eta}\,,\\
\nonumber \\ 
c_2 & = \frac{5 \pi}{32514048} \frac{1}{\eta^{5} M^5} \bigg [(m_1^5 s_2^2 + m_2^5 s_1^2)(-4341025 + 65553264 \eta - 684432 \eta^2)
\nonumber \\
\nonumber \\
    & \quad + \eta M^2 ( m_1^3 s_2^2 + m_2^3 s_1^2) (20044511 + 65553264 \eta - 
   684432 \eta^2)
\nonumber \\
   & \quad +42 \eta^2 M^5 s_1 s_2 (1029619 - 36387504 \eta - 7970256 \eta^2) \bigg ]
-\frac{15  f_4^\GB}{64 \eta}. 
\end{align}
\end{widetext}
Here $\eta \equiv m_1 m_2/M^2$ is the symmetric mass ratio while $f_3^\GB$ and $f_4^\GB$ represent our ignorance of the correction to the tensor non-dipole emission in EdGB gravity at 1.5PN and 2PN orders\footnote{We have replaced $f_i^\ST$ in \cite{EdGB2016wfm} to $(\alpha_\GB^2/M^4) f_i^\GB$ for $i=3,4$.}. The above corrections can be mapped to the parameterized post-Einsteinian (PPE) framework \cite{ppE2009,ppE2012,ppE2018} of 
\begin{equation}
\delta \Psi = \sum_i \beta_i^\PPE v^{-5+2i}\,,
\end{equation}
with
\begin{equation}
\beta_i^\PPE = \frac{\alpha_\GB^2}{M^4} c_i\,.
\end{equation}
The leading $-1$PN term ($c_{-1}$ or $\beta^\PPE_{-1}$) derived here agrees with those found in \cite{Yagi:2011xp,Yunes:2016jcc}. 

\begin{figure}
	\centering
    \includegraphics[width=0.43\textwidth]{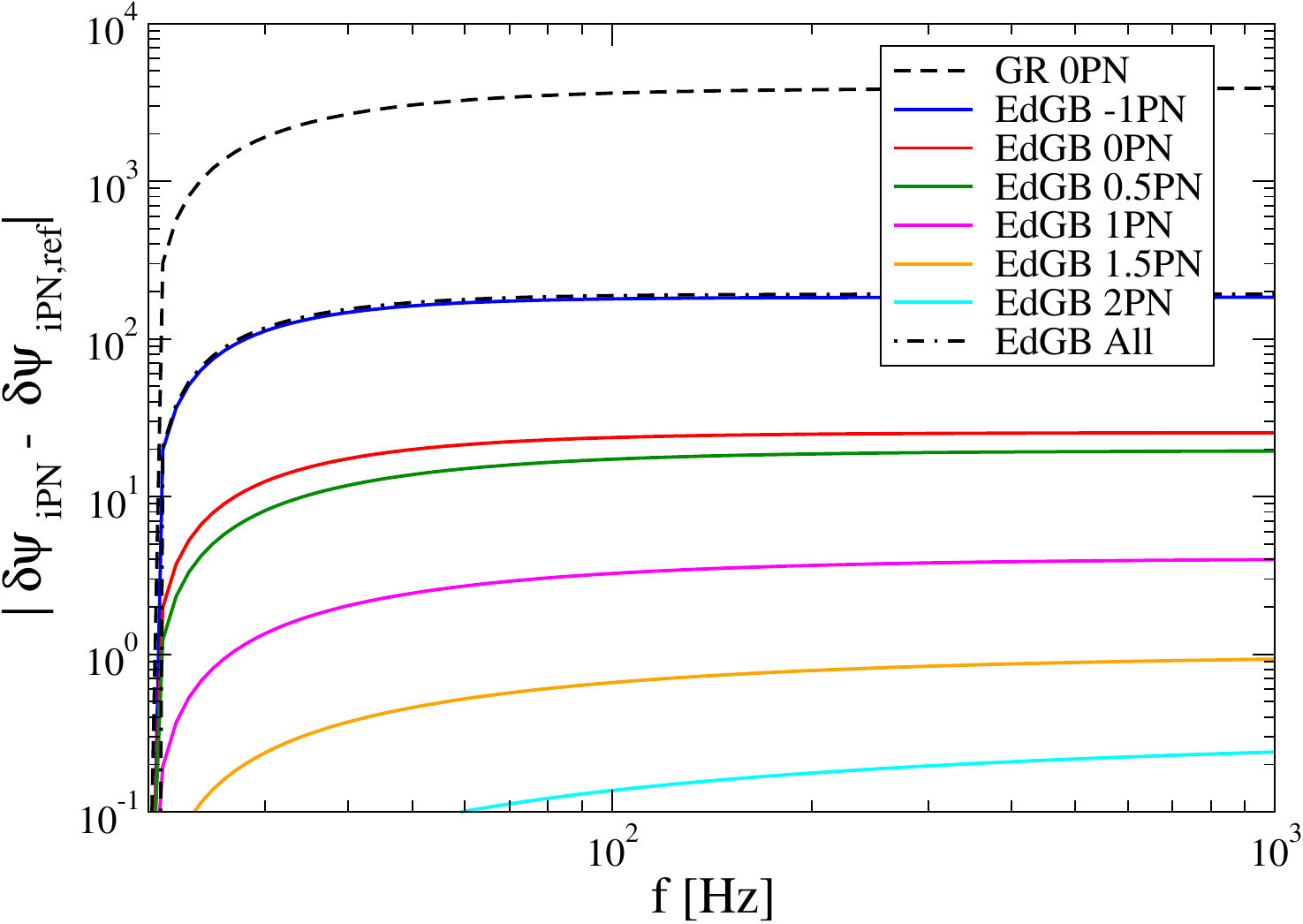}
	\caption{
	    Comparison of EdGB corrections to the phase entering at different PN orders as a function of the GW frequency. We also present the leading phase in GR and the contribution from all of the EdGB corrections combined. For each contribution, we show the phase relative to that at a reference frequency chosen to be 20Hz. We chose $(m_1,m_2) = (5.9,1.4) M_\odot$ and $(\chi_1,\chi_2)=(0.31,0)$, corresponding to GW200115, and $\sqrt{\alpha_\GB} = 1.33\,\rm km$ that is the 90\% credible limit found through our Bayesian inference (see Table \ref{tab:result}).
} 
	\label{fig:phase}
\end{figure}

Figure \ref{fig:phase} presents each PN correction term in the phase against the GW frequency $f$ for GW200115, together with the leading GR term. We chose $\sqrt{\alpha_\GB} = 1.33$km that is the 90\% credible limit found through our Bayesian inference in Table \ref{tab:result}. Notice that the EdGB corrections are subdominant to GR by at least an order of magnitude. Notice also that the leading EdGB correction at $-1$PN order dominates higher PN contributions at $f \lesssim 200$Hz and the latter becomes only important when the frequency becomes high (though the noise becomes larger as the frequency becomes higher), which explains why higher PN corrections do not affect the bound on $\sqrt{\alpha_\GB}$ much. It is interesting to note that for $f \gtrsim 200$Hz, the EdGB phase is dominated by the contribution at 1.5PN order, though the phase is still incomplete at this order (we have set the unknown contributions $f_3^\GB$ and $f_4^\GB$ to 0 in Fig. \ref{fig:phase}). 

Let us comment on up to which PN order the above waveform corrections are complete. The $\alpha_\GB$ dependence in the above corrections enter only through the scalar charges $\alpha_A$. There are other contributions to the waveform where $\alpha_\GB$ appears explicitly though such contributions enter at 3PN order and are negligible for our purpose \footnote{We count the PN order in powers of $v/c$ while Shilarirou et al. \cite{EdGB2021wfm,EdGB2021wfm02} counts in powers of $1/c$. With the latter counting, the $\alpha_\GB$ dependence other than scalar charges enters at 1PN.}. For non-spinning binaries, they are complete up to 1PN order. The expressions at 1.5PN and 2PN include currently unknown $f_3^\GB$ and $f_4^\GB$ but they also have other missing contributions, such as the scalar dipole radiation at 1.5PN and 2PN orders (which correspond to 2.5PN and 3PN relative to the leading $-1$PN contribution) and the correction to the binding energy or Kepler's law at 3PN that couples to the $-1$PN dipole radiation and enter at 2PN in the waveform. For spinning binaries, the waveform is complete only up to 0PN order as the effect of spins are only included through the scalar charges $\alpha_A$. Missing contributions include e.g. a spin-orbital coupling in the binding energy at 1.5PN order that couples with the leading dipole radiation to enter at 0.5PN in the waveform.\\

We end by comparing the 0PN corrections found here with different functional forms considered in  \cite{EdGB_Perkins}. Using Eqs. \eqref{eq:cm1} and \eqref{eq:c0}, the 0PN correction to the phase can be expressed as:
\begin{equation}
\label{eq:Psi_0PN}
\delta \Psi_\mathrm{0PN} =  \frac{659 + 728 \eta}{96} v^2 \delta \Psi_\mathrm{-1PN} + \frac{5 \pi}{16}\frac{s_1 s_2}{\eta ^3}\frac{\alpha_\GB^2}{M^4} v^{-5}\,.
\end{equation} 
The first term is similar to one of the functional forms considered in \cite{EdGB_Perkins}:
\begin{equation}
\delta \Psi_\mathrm{0PN}^\mathrm{(PNSY,1)} = \frac{5}{756} (743+924 \eta) \gamma \, u^2 \, \delta \Psi_\mathrm{-1PN}\,,
\end{equation}
where $\gamma$ is a constant that does not depend on binary parameters, $u \equiv (\pi \mathcal M f)^{1/3}$ and the $\eta$ dependence is taken from that in the phase at 1PN order in GR. The $\eta$ dependence in the two expressions,however, are different. The second term in Eq. \eqref{eq:Psi_0PN} is similar to another functional form considered in \cite{EdGB_Perkins}:
\begin{equation}
\label{eq:Perkins}
\Psi_\mathrm{0PN}^\mathrm{(PNSY,2)} = 16\pi \frac{\alpha_\GB^2}{M^4} \gamma u^{-5}\,,
\end{equation}
though again, the expressions are different. This is because if one maps the second term in Eq. \eqref{eq:Psi_0PN} to Eq. \eqref{eq:Perkins}, $\gamma$ depends on binary parameters through $\eta$ and $s_A$.

\bibliography{ref.bib}

\end{document}